# Demagnetization dynamics after noncollinear dual optical excitation

Sergii Parchenko[1]*, Matthias Riepp[2,3], Simon Marotzke[4,5], Agne Åberg Larsson[6], Vassilios Kapaklis[6], Peter M. Oppeneer[6], Andreas Scherz[1]

1. European XFEL, Holzkoppel 4, 22869 Schenefeld, Germany
2. Université de Strasbourg, CNRS, Institut de Physique et Chimie des Matériaux de Strasbourg, UMR 7504, 67000 Strasbourg, France
3. Sorbonne Université, CNRS, Laboratoire de Chimie Physique – Matière et Rayonnement, LCPMR, 75005 Paris, France
4. Deutsches Elektronen-Synchrotron DESY, Notkestr. 85, 22607 Hamburg, Germany
5. Institut für Experimentelle und Angewandte Physik, Christian-Albrechts-Universität zu Kiel, 24098 Kiel, Germany
6. Department of Physics and Astronomy, Uppsala University, Box 516, SE-75120 Uppsala, Sweden

*Corresponding author: sergii.parchenko@xfel.eu

*Abstract.* We explore the impact of optical excitation using two interfering ultrashort optical pulses on ultrafast magnetization dynamics. Our investigation focuses on Pt/Co/Pt multilayers and TbCo alloy samples, employing a dual pump approach. We observe significant variations in the dynamics of magnetization suppression and subsequent recovery when triggered with two optical pulses of the same polarization—essentially meeting conditions for interference. Conversely, dynamics triggered with cross-polarized pump beams exhibit expected similarity to that triggered with a single pulse. Delving into the underlying physical processes contributing to laser-induced demagnetization and recovery dynamics, we find that our current understanding cannot elucidate the observed trends. Consequently, we propose that optical excitation with interfering light possesses not previously acknowledged capacity to induce long-lasting alterations in the dynamics of angular momentum.

It was demonstrated recently that ultrafast excitation with two interfering noncollinear optical pulses induces magnetization dynamics significantly distinct from that triggered by a single excitation pulse [1]. Specifically, the decay time, frequency, and amplitude of magnetization precession in a permalloy film following dual optical excitation exhibit a trend that cannot be explained within the classic understanding of the interaction between ultrashort laser pulses and a magnetic medium. Certain consistencies with the experimental results were found when considering the action of the opto-magnetic field from pump pulses on the magnetic moments. However, the explanation of a time window of the effect, spanning almost a nanosecond, remained unclear. Consequently, the influence of a previously unknown opto-magnetic effect was hypothesized to be a potential cause of the observed results. Namely, it was hypothesized that the optical excitation with two noncollinear ultrashort laser pulses that interfere alters the magnetic properties of the material, and the induced modification persists for a time interval that significantly exceeds the duration of the two ultrashort laser pulses. If the suggested effect does exist, (i) it should be observed in different magnetic materials, and (ii) other types of ultrafast magnetization dynamics should also deviate from the single-pulse excitation regime.

Here, we present our investigation of ultrafast demagnetization dynamics after noncollinear dual optical excitation. We examine two different systems that have been widely studied before with optical time-resolved methods: the Pt/Co/Pt multilayer and the TbCo alloy. In both samples, we observe a clear difference between the demagnetization dynamics induced by two interfering optical pump pulses and the dynamics induced by two noninterfering pump pulses or the classic single-pulse excitation regime. We show that the observed difference could not originate from any form of spatial variation in excitation efficiency associated with creation of transient gratings. Our observations further support the hypothesis that optical excitation with interfering light can cause long-lasting modification of the dynamic magnetic behavior of the materials.

Since its first demonstration almost three decades ago [2], ultrafast laser-induced demagnetization dynamics have been the subject of extensive experimental and theoretical investigations [3, 4]. The interest in laser-induced demagnetization is driven by a combination of interest in the fundamental processes involved in the phenomenon and potential technological applications in next-generation data processing and storage devices that utilize the principle of optical control of magnetization [5, 6]. Significant achievements have been made in revealing the underlying mechanisms. Most of the current understanding of the ultrafast demagnetization process relies on the process of angular momentum transfer from optically excited electrons to other subsystems of the material. The common argument, based on the phenomenological three-temperature model (3TM) [2], is that hot electrons transfer angular momentum, associated with magnetization, to the lattice and the spin system. The subpicosecond timescale of demagnetization and recovery dynamics is determined by the electron-electron, electron-spin, and electron-lattice coupling times. Several possible dissipation channels have been discussed including electron-lattice spin-flip scattering [7, 8] and electron-magnon interaction [9, 10, 11]. Also, an alternative approach involves spin transport across the sample [12, 13, 14]. The subsequent rapid partial recovery of magnetization is driven by the same principles, but instead of transmitting angular momentum to the lattice, electrons gain angular momentum from the lattice. Even though the contribution of particular scattering mechanisms varies in different systems and the relative contribution is still the subject of debate, within the 3TM the dynamics is nonetheless expected to be the same when triggered either with a single pulse or with two overlapping ultrashort laser pulses that come from different directions. For an optically isotropic magnetic media, the triggered ultrafast demagnetization and recovery dynamics must be the function of the number of absorbed photons and any additional non-linear absorption effects that could alter the ultrafast magnetization dynamics should show a similar trend in the case of single or dual pulse excitation approaches.

In our study, we investigated two systems: a Pt/Co/Pt multilayer and a TbCo alloy. The Pt/Co/Pt multilayer sample was fabricated using electron-cyclotron (ECS) and direct current (DC) magnetron sputtering on a $Si_3N_4$ substrate. The specific sample structure was Pt(2 nm)/[Co(0.8 nm)/Pt(1.4 nm)]$_{x8}$/Pt(2 nm DC)/Pt(4 nm ECS)/$Si_3N_4$. This Pt/Co/Pt configuration exhibits out-of-plane anisotropy with a coercive field $H_C$ of approximately 25 mT and a saturation field $H_S$ of approximately 250 mT. For the TbCo alloy, an amorphous film with a composition of $Tb_{18}Co_{22}$ and a thickness of 20 nm was prepared using DC magnetron sputtering. The resulting sample structure comprises an $Al_2O_3$ capping layer (2 nm)/$Tb_{18}Co_{22}$ film (20 nm)/$Al_2O_3$ seed layer (2 nm)/$SiO_2$ substrate. More detailed information about the TbCo film can be found in [15]. Like for the Pt/Co/Pt multilayer, the TbCo film also exhibited out-of-plane magnetic anisotropy with a coercive field $H_C$=250 mT.

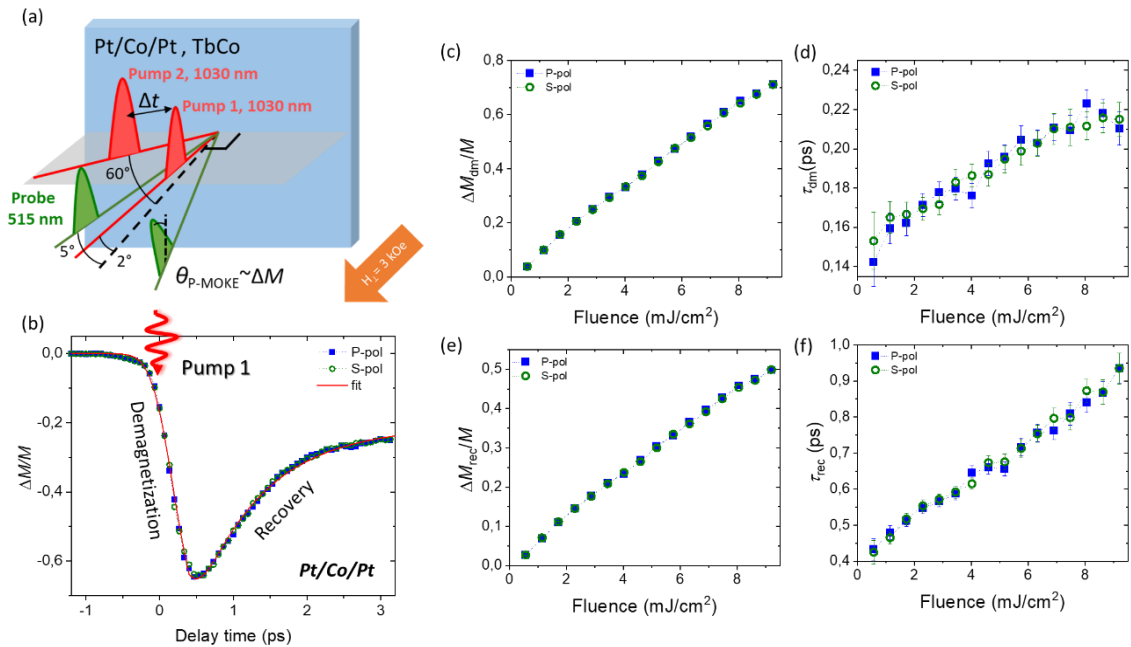

Figure 1. (a) Experimental geometry of dual optical excitation experiments. (b) Time-resolved demagnetization dynamics in Pt/Co/Pt after excitation only with S- or P-polarized pump 1 with fluence F = 7.5 mJ/cm$^2$. The red line is fit to eq. 1. (c) Demagnetization amplitude $\Delta M_{dm}$, (d) demagnetization time $\tau_{dm}$, (e) recovery amplitude $\Delta M_{rec}$, and (f) recovery time $\tau_{rec}$ as a function of pump 1 fluence for P and S polarization. The error bars are the 95% confidence intervals. For panels (c) and (e), the error bars are smaller than the size of the data markers.

The experimental geometry for dual-pump time-resolved experiments closely resembles the setup used in [1] and is depicted in Fig. 1(a). Ultrashort laser pulses were generated by a Yb-fiber laser with a fundamental wavelength $\lambda$ = 1030 nm and a repetition rate of 50 kHz. Optical excitation involved the application of two linearly polarized pulses, each lasting 300 fs, and characterized by a wavelength of $\lambda$ = 1030 nm. The pump spot size was set to 250 μm, with incidence angles of 2° and 60° for pump 1 and pump 2, respectively. The polarization direction of pump 1 was varied using a half-wave plate, while that of pump 2 remained P-polarized throughout all the experiments. Linearly polarized optical probe pulses with a wavelength of $\lambda$ = 515 nm and a spot size of 100 µm were generated through frequency doubling of the laser's fundamental wavelength. The probe beam was incident at an angle of approximately 5° to the surface normal. The probe for transient magnetization change $\Delta M$ was based on analyzing the Kerr rotation of the polarization plane of the reflected probe beam using a balanced detection scheme. In this arrangement, the primary magnetic signal was derived from the polar magneto-optical Kerr effect (MOKE), which is sensitive to the out-of-plane component of magnetization. The pump-probe delay time presented in the plots below represents the separation between pump 1 and probe pulses. An external magnetic field of $H$ = 300 mT was applied perpendicular to the sample surface, sufficient to maintain the monodomain magnetic state in both samples.

Figure 1(b) shows the transient magnetization change $\Delta M/M$ in the Pt/Co/Pt sample after excitation only with pump 1 with fluence $F$=7.5 mJ/cm$^2$. Following the rapid decrease in the $\Delta M/M$ signal on a subpicosecond timescale, corresponding to demagnetization dynamics, partial recovery of $\Delta M/M$ is observed a few picoseconds after excitation. The observed dynamics align with previously reported observations [16, 17]. Traces taken under P- and S-polarized pump pulses show no difference, as expected for the thermal mechanism of excitation and equal absorption for P- and S-polarized light at nearly perpendicular incidence. Two exponential functions with convolution to the pulse duration were used to fit the time traces:

$$\Delta M(t)/M = \Delta M_{dm}/M \cdot e^{-t/\tau_{dm}} + \Delta M_{rec}/M \cdot e^{-t/\tau_{rec}} \qquad (1)$$

where $\Delta M_{dm(rec)}$ and $\tau_{dm(rec)}$ represent the amplitude and characteristic time of the demagnetization and recovery components, respectively. The panels in Fig. 1(c), (d), (e), and (f) show the values of $\Delta M_{dm}$, $\tau_{dm}$, $\Delta M_{rec}$, and $\tau_{rec}$, respectively, as a function of the pump 1 fluence. All four quantities linearly increase with increasing optical excitation fluence, what is consistent with previously reported observations [8]. It should be noted that the demagnetization time of Co is on the order of 100 fs [17], which is shorter than the pulse duration in our experiments, resulting in a situation when demagnetization takes place when the excitation pulse is present in the material, thereby smearing out the dynamics.

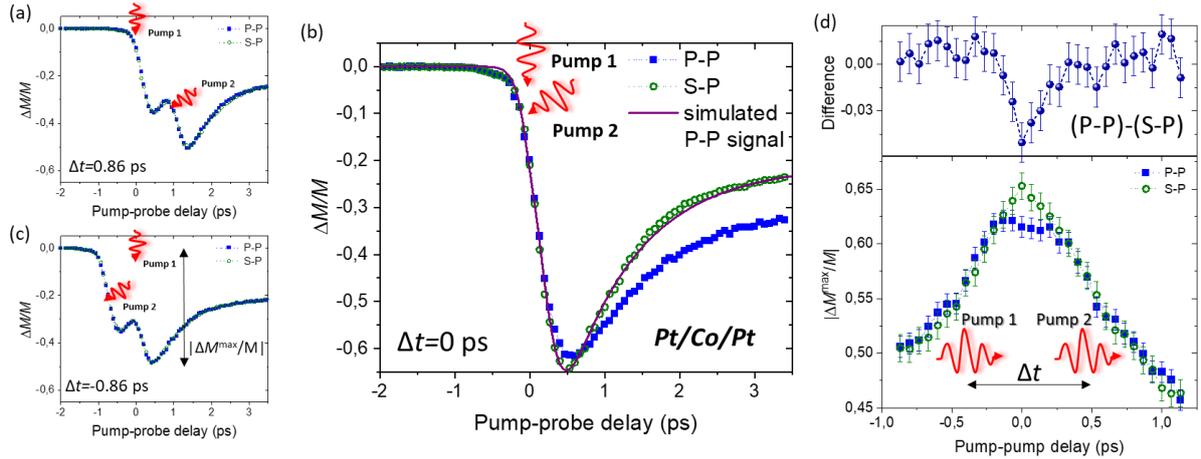

*Figure 2. (a), (b) and (c) Demagnetization dynamics in Pt/Co/Pt after excitation with two pump pulses with an equal fluence of 4 mJ/cm$^2$ for pump-pump delay times Δt = 0.86 ps, 0 ps and -0.86 ps, respectively. Pump 1 was either P- or S-polarized, and pump 2 was always P-polarized. The purple line in panel (b) is the simulated ΔM/M signal for P-P polarizations considering only the periodic alternation of the excitation strengths. See the main text for more details. (d) The absolute value of the maximum magnetization suppression |ΔM$^{max}$/M| as a function of pump-pump delay Δt for the P-P and S-P configurations (bottom panel) and the difference between the two dependencies (top). The inset to the bottom panel describes the determination of Δt.*

When employing two optical pulses to induce the demagnetization dynamics, notable distinctions arise when compared to the single-pulse excitation regime. The fluence of pump 1 and pump 2 was $F$=4 mJ/cm$^2$. Panels (a), (b), and (c) of Fig. 2 illustrate the magnetization dynamics in Pt/Co/Pt following excitation with pump 1 and pump 2 for different pump-pump delay times Δ$t$. In instances where the two pump pulses are separated by a time window exceeding the pulse duration, there is no discernible difference in the dynamics when pump 1 is either P- or S-polarized, denoted as the P-P and S-P configurations, respectively (with pump 2 always being P-polarized). However, a clear difference in magnetization dynamics is evident when pump-pump delay time Δ$t$ = 0 ps (see Fig. 2(b)). When two noncollinear optical pulses with the same polarization overlap in time, they are in interference conditions. This interference induces a spatially periodic modulation of excitation efficiency: the sample is more strongly pumped at interference maxima and less pumped at minima. Considering the linear fluence dependence of all four parameters that describe the magnetization dynamics (refer to Fig. 1), it is possible to simulate the expected Δ$M$(t)/$M$ signal for optical excitation with interfering light considering a periodic excitation pattern. This simulation takes into account only the variation in excitation efficiency while assuming that the material behavior remains the same (refer to the Supplementary Materials for more information). The simulated signal, depicted in Fig. 2(b) by the purple line, closely resembles the time trace measured in the S-P configuration. Consequently, the observed discrepancy in magnetization dynamics cannot be solely explained by considering periodic excitation patterns and must arise from a light-induced modification of the magnetic properties of the material. Consistent with the results reported in [1], the difference in dynamics is evident only within the Δ$t$ window, which is compatible with the pump pulse duration in our experiments (see Fig. 2(d)).

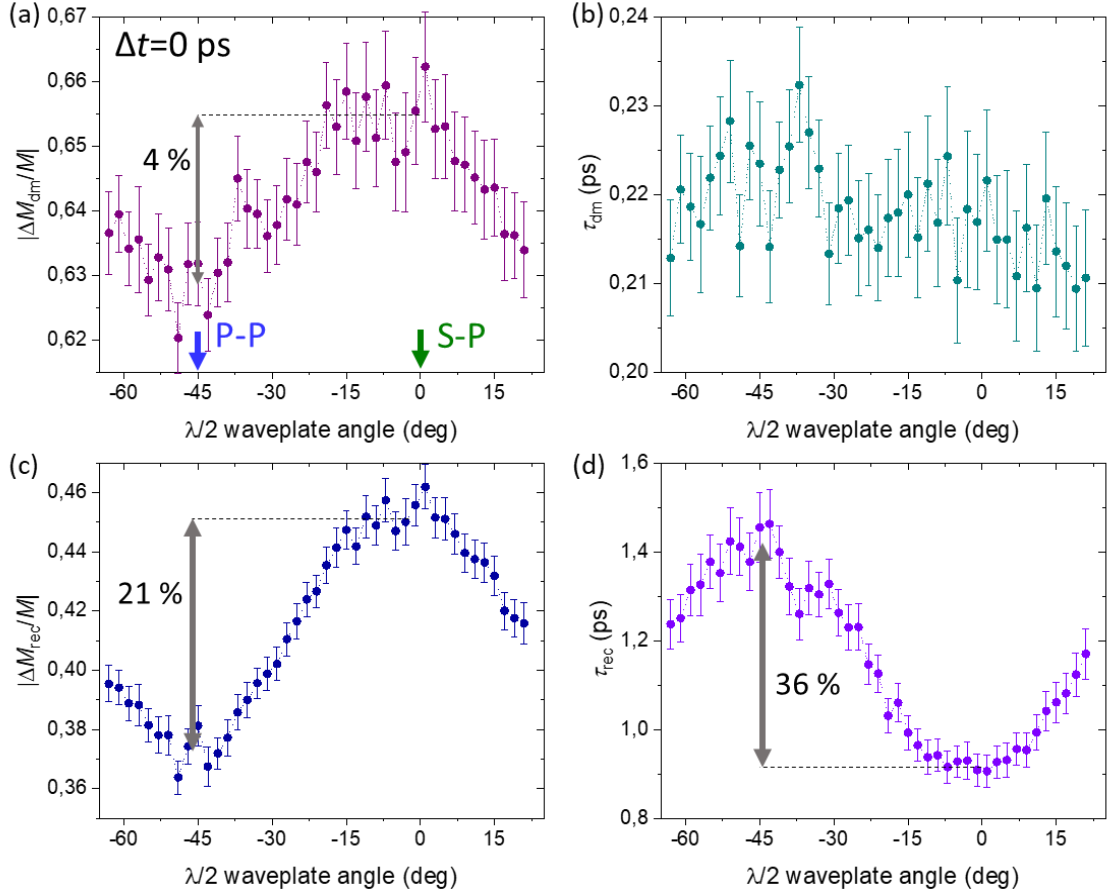

Figure 3. Fit results of dynamics in Pt/Co/Pt: (a) Demagnetization amplitude Δ$M_{dm}$, (b) demagnetization time $τ_{dm}$, (c) recovery amplitude Δ$M_{rec}$, and (d) recovery time $τ_{rec}$ as a function of the pump 1 polarization direction, determined by the rotation of the half-wave plate, after excitation with two pump beams. Pump 2 was always P-polarized. The fluence for the two pumps was 4 mJ/cm². The half-wave plate position when excited in the P-P and S-P configurations is indicated in panel (a).

Furthermore, we delve into the magnetization dynamics after dual excitation for pump-pump delay Δ$t$=0 ps, exploring the variation in interference efficiency by rotating the polarization direction of pump 1 in Pt/Co/Pt. The results are summarized in Fig. 3, where all four graphs depict the variation in the respective parameters as a function of the polarization direction of pump 1. The demagnetization Δ$M_{dm}$ and recovery Δ$M_{rec}$ amplitudes reach their maximum when the two pumps are out of interference conditions (S-P configuration) and are minimized when interference conditions are perfectly met. However, the relative differences in these values are not the same (refer to Fig. 3).

Furthermore, given the linear increasing trend of $\Delta M_{dm}$, $\Delta M_{rec}$, $\tau_{dm}$, and $\tau_{rec}$ as a function of excitation strength (refer to Fig. 1), one would expect all four quantities to follow the same tendency. Simultaneously, the dependence of the demagnetization time $\tau_{dm}$ and recovery time $\tau_{rec}$ shows the opposite trend. The time constants are largest when interference occurs between the two pumps and smallest when interference is suppressed. While the change in demagnetization time is only slightly above the errorbars, the change in recovery time is very evident.

During the demagnetization process, induced by ultrashort laser pulses, there is a transfer of angular momentum from electrons to the lattice. Conversely, during the recovery phase, this transfer of angular momentum occurs in the opposite direction. The most widely discussed mechanism driving ultrafast demagnetization and recovery dynamics is electron-phonon spin-flip scattering, as described by the Elliott–Yafet mechanism [3]. In the demagnetization stage, an optically excited electron can change its spin by emitting a phonon or magnon. This process takes place rapidly, within a sub-picosecond timescale, particularly when electrons exist in an laser induced out-of-equilibrium state. Subsequently, during the fast recovery dynamics, which occur on a picosecond timescale, the lattice returns a portion of the previously transferred angular momentum to the electrons. The spin-flipping process is facilitated by the absorption of phonons and is influenced by the electron-lattice relaxation time. For a given material, the Elliott–Yafet spin-flip probability is a constant value related only to the spin-orbit coupling and exchange splitting and thus to the magnetic moment of the atom [25,26,27]. The timescale and magnitude of both the demagnetization and recovery phases should increase linearly with increasing fluence of optical excitation, as shown also in our experiments with a single excitation pulse. Next, in our experiments with two excitation pulses, the efficiency of optical excitation within the probed area is the same for the P-P and S-P configurations. Pump 1, with varying polarization, impinges on the material almost at normal incidence, and the polarization of pump 2 is constant. Thus, considering nearly the same absorbed fluence, the dynamics in P-P and S-P configurations must be identical. However, the experimental observations clearly show the opposite. It should be also noted that any non-linear absorption effects that might cause changes in the magnetization dynamics should be seen also in single-pulse excitation experiments. The observed dynamics after dual pulse excitation in the S-P configuration is very similar to that after a single pulse excitation as can be seen from the comparison of Fig. 1(b) and Fig. 2(b). The magnetization response is different only when two pump pulses are in interference. Thus, the observed variations in the magnitude and timescales of demagnetization and recovery dynamics, together with previously reported findings about the magnetization precession, further indicate that ultrafast optical excitation with two noncolinear interfering pulses changes the magnetic behavior of the materials related to the dynamics of angular momentum.

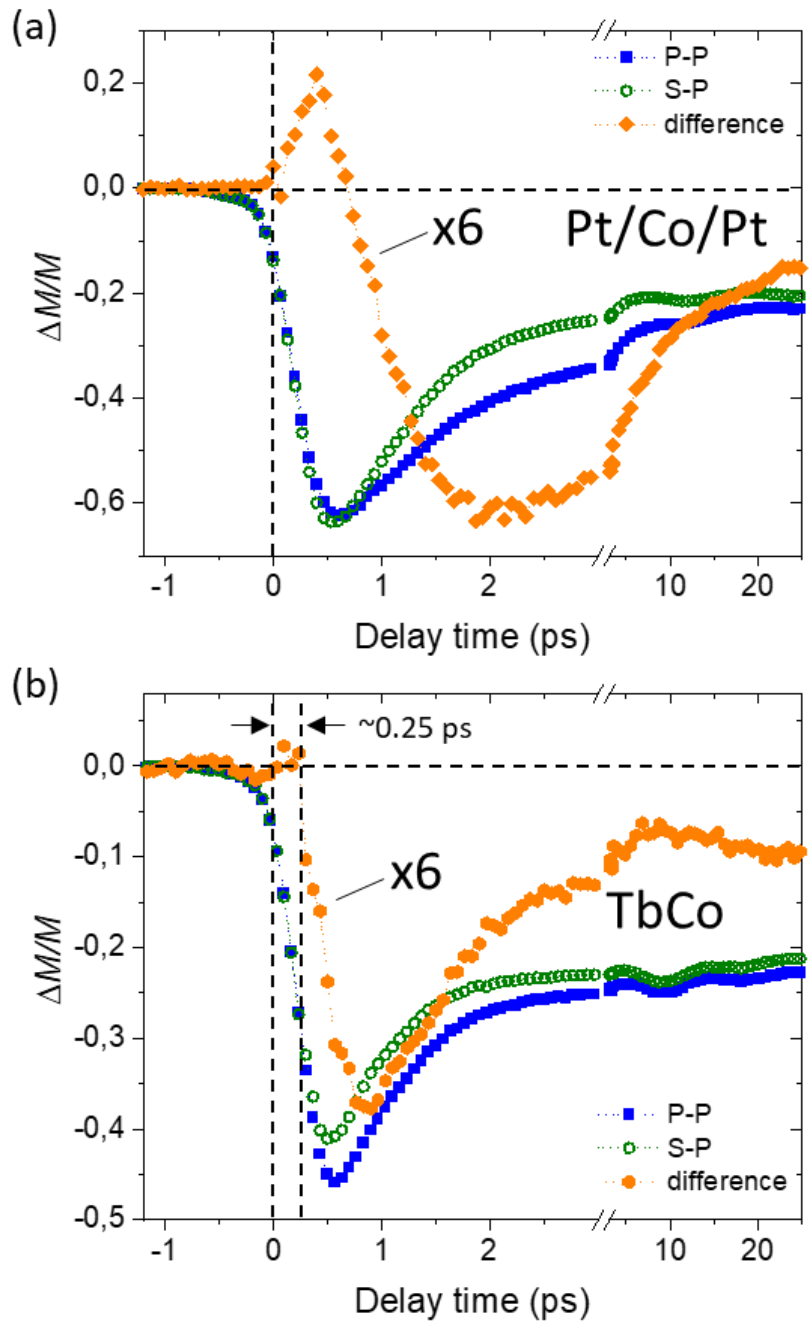

*Figure 4. (a) Demagnetization dynamics in Pt/Co/Pt multilayer and (b) TbCo alloy shown for an extended delay range for P-P and S-P polarizations. The orange trace is the difference between curves taken with different polarizations. The pump-pump delay time is Δt=0 ps.*

The impact of optical excitation with interfering ultrashort light pulses on the dynamics in TbCo alloy exhibits noteworthy differences compared to that in the Pt/Co/Pt system, as illustrated in Fig. 4, where demagnetization dynamics for both samples are depicted for an extended delay time. The optical fluence of both pump beams when exciting the TbCo sample was $F$=1.5 mJ/cm$^2$. The TbCo displays a more pronounced suppression of magnetization when two pulses interfere while for Pt/Co/Pt magnetization suppression was smallest in the P-P configuration. The extracted parameters ($\Delta M_{dm}$, $\Delta M_{rec}$, $\tau_{dm}$, and $\tau_{rec}$) for TbCo as a function of pump 1 polarization are detailed in the Supplementary Materials. The orange curves in both panels of Fig. 4 represent the difference between time traces collected after optical excitation with temporarily overlapping pump pulses in the P-P and S-P configurations. Notably, the difference in dynamics between the P-P and S-P configurations in Pt/Co/Pt is evident from the initial moment of excitation, while for TbCo, the initial 0.25 ps of dynamics are the same. Furthermore, the disparate dynamics in Pt/Co/Pt persist for more than 20 ps after the excitation, while there is almost no difference in TbCo already after 3 ps.

The observed delay in the reaction to optical excitation with interfering pulses for TbCo may indicate that the possible long-lasting modification of angular momentum transfer could be induced in electrons that were not directly affected by optical excitation with interfering photons. In Pt/Co/Pt we optically excite and track the change in magnetization of Co, which is the dominant magnetic element in the system. Induced magnetization of Pt due to proximity effects we consider to have a minor contribution to the measured signal. However, for TbCo, NIR optical excitation initially affects delocalized 3$d$ electrons of Co and 5$d$ electrons of Tb [8]. Later, the excitation is transferred from 5$d$ electrons to more localized 4$f$ electrons of Tb [9]. With $\lambda_{probe}$=515 nm, the probed signal contains a mixture of magnetic moments of 4$f$ and 5$d$ electrons of Tb ions, with the 4$f$ contribution expected to be dominant due to the significantly higher magnetic moment associated with 4$f$ states [18, 19, 20]. However, there are some contributions from 5$d$ states in the probed signal. If the modification of electron properties, associated with dual pump excitation, would affect only the optically excited 5$d$ electrons of Tb, we would expect to see a difference in magnetization dynamics between the P-P and S-P configurations that is proportional only to the difference in the demagnetization and recovery rate of 5$d$ electrons and should be seen from the initial moment of excitation. In addition, the shorter period for TbCo where the dynamics between S-P and P-P excitation are disparate may suggest that the recovery of electrons to the ordinary state occurs faster in more localized 4$f$ states than in delocalized 3$d$ states. Thus, the delay in the difference between magnetization dynamics in P-P and S-P configurations in TbCo and the different temporal evolution of disparity between dynamics in P-P and S-P configurations is an indication that the properties of 4$f$ electrons were affected indirectly via interactions with electrons that were initially modified by dual pump excitation. These considerations provide valuable insights into describing of modification of magnetization dynamics after dual-pump excitation but require further investigation and can be confirmed only by the clear separation of 4$f$ and 5$d$ dynamics [9], which is challenging with optical methods.

Finally, we would like to clarify that the idea of using two optical pulses in interference for ultrafast excitation is not particularly new. Several past works reported on magnetization dynamics after excitation with two optical pulses, where the main aspect was on creation of transient gratings [21, 22, 23, 24]. However, the separation of effects related to periodic excitation patterns, reported previously, and modification of magnetization dynamics, described in this study, requires a careful comparison of dynamics excited when two pump pulses are in and not in interference, which was not done before.

In summary, our study provides further evidence that optical excitation with two noncollinear ultrashort pulses that interfere induces long-lasting modification of the magnetization dynamics. This modification extends beyond the previously reported effects on magnetization precession to influence optically induced demagnetization and recovery dynamics. Examining the mechanisms responsible for these processes we suggest that the electron angular momentum dynamics might be altered due to the optical excitation with interfering light. If confirmed, this might potentially imply the possibility of altering the fundamental properties of electrons, particularly that related to angular momentum, which is currently considered constant for all electrons. Modification of electron behavior with dual interferometric optical excitation introduces new possibilities in physics. The angular momentum of electrons, regarded as a fundamental property of their wavefunction with no analogy in the macroscopic world, remains a subject of ongoing discussion and investigation. At this stage, providing a definitive explanation for the observed effect is challenging, given the unknown nature of the angular momentum of elementary particles. However, this discovery opens new horizons in physics, and further exploration of this effect, especially with laser fluences significantly exceeding those used in the current study, may lead to breakthroughs in our understanding of fundamental physical phenomena beyond current knowledge boundaries.

**Acknowledgments:**

PMO acknowledges financial support from the Swedish Research Council (VR, Grants No. 2021-05211 and 2022-06725) and the Knut and Alice Wallenberg Foundation (Grants No. 2022.0079 and 2023.0336). MR was supported on the ANR-20-CE42-0012-01 (MEDYNA) grant.

**References**

1. Sergii Parchenko, Davide Pecchio, Ritwik Mondal, Peter M. Oppeneer, Andreas Scherz, Magnetization precession after non-collinear dual optical excitation, https://doi.org/10.48550/arXiv.2305.00259
2. Beaurepaire, E., Merle, J.-C., Daunois, A. & Bigot, J.-Y. Ultrafast Spin Dynamics in Ferromagnetic Nickel. *Phys. Rev. Lett.* **76**, 4250–4253 (1996).

3. Kirilyuk, A., Kimel, A. V. & Rasing, T. Ultrafast optical manipulation of magnetic order. *Rev. Mod. Phys.* **82**, 2731 (2010).

4. Scheid, P., Remy, Q., Lebègue, S., Malinowski, G. & Mangin, S. Light induced ultrafast magnetization dynamics in metallic compounds. *J. Magn. Magn. Mat.* **560**, 169596 (2022).

5. Kimel, A. V. & Li, M. Writing magnetic memory with ultrashort light pulses. *Nat. Rev. Mater.* **4**, 189–200 (2019).

6. El-Ghazaly, A., Gorchon, J., Wilson, R. B., Pattabi, A. & Bokor, J. Progress towards ultrafast spintronics applications. *J. Magn. Magn. Mat.* **502**, 166478 (2020).

7. Koopmans, B., Ruigrok, J. J. M., Longa, F. D. & de Jonge, W. J. M. Unifying Ultrafast Magnetization Dynamics. *Phys. Rev. Lett.* **95**, 267207 (2005).

8. Koopmans, B. *et al.* Explaining the paradoxical diversity of ultrafast laser-induced demagnetization. *Nat. Mat.* **9**, 259–265 (2010).

9. Frietsch, B. *et al.* The role of ultrafast magnon generation in the magnetization dynamics of rare-earth metals. *Sci. Adv.* **6**, eabb1601 (2020).

10. Haag, M., Illg, C. & Fähnle, M. Role of electron-magnon scatterings in ultrafast demagnetization. *Phys. Rev. B* **90**, 014417 (2014).

11. Weißenhofer, M. & Oppeneer, P. M. Ultrafast Demagnetization Through Femtosecond Generation of Non-Thermal Magnons. *Adv. Physics Res.*, 2300103 (2024).

12. Battiato, M., Carva, K. & Oppeneer, P. M. Superdiffusive Spin Transport as a Mechanism of Ultrafast Demagnetization. *Phys. Rev. Lett.* **105**, 027203 (2010).

13. Eschenlohr, A. *et al.* Ultrafast spin transport as key to femtosecond demagnetization. *Nat. Mat.* **12**, 332–336 (2013).

14. Razdolski, I. *et al.* Nanoscale interface confinement of ultrafast spin transfer torque driving non-uniform spin dynamics. *Nat. Comm* **8**, 15007 (2017).

15. Ciuciulkaite, A. *et al.* Magnetic and all-optical switching properties of amorphous Tb x Co 100 − x alloys. *Phys. Rev. Mat.* **4**, (2020).

16. Cardin, V. *et al.* Wavelength scaling of ultrafast demagnetization in Co/Pt multilayers. *Phys. Rev. B* **101**, 054430 (2020).

17. Vaskivskyi, I. *et al.* Element-Specific Magnetization Dynamics in Co–Pt Alloys Induced by Strong Optical Excitation. *J. Phys. Chem. C* **125**, 11714–11721 (2021).

18. Alebrand, S. *et al.* Subpicosecond magnetization dynamics in TbCo alloys. *Phys. Rev. B* **89**, 144404 (2014).

19. Khorsand, A. R. *et al.* Element-Specific Probing of Ultrafast Spin Dynamics in Multisublattice Magnets with Visible Light. *Phys. Rev. Lett.* **110**, (2013).

20. Chen, Z., Li, S., Zhou, S. & Lai, T. Ultrafast dynamics of 4f electron spins in TbFeCo film driven by inter-atomic 3d–5d–4f exchange coupling. *New J. Phys.* **21**, 123007 (2019).

21. Ksenzov, D. *et al.* Nanoscale Transient Magnetization Gratings Created and Probed by Femtosecond Extreme Ultraviolet Pulses. *Nano Lett.* **21**, 2905–2911 (2021).

22. Frazer, T. D. *et al.* Optical transient grating pumped X-ray diffraction microscopy for studying mesoscale structural dynamics. *Sci. Rep.* **11**, 19322 (2021).

23. Janušonis, J. *et al.* Transient Grating Spectroscopy in Magnetic Thin Films: Simultaneous Detection of Elastic and Magnetic Dynamics. *Sci. Rep.* **6**, 29143 (2016).

24. Wang, G. *et al.* Gate control of the electron spin-diffusion length in semiconductor quantum wells. *Nat. Comm.* **4**, 2372 (2013).

25. Steiauf, D. and Fähnle, M., Elliott-Yafet mechanism and the discussion of femtosecond magnetization dynamics. *Phys. Rev. B* **79**, 140401(R) (2009).

26. Carva, K., Battiato, M., and Oppeneer, P. M., Ab Initio Investigation of the Elliott-Yafet Electron-Phonon Mechanism in Laser-Induced Ultrafast Demagnetization. *Phys. Rev. Lett.* **107**, 207201 (2011).

27. Essert, S. and Schneider, H. C., Electron-phonon scattering dynamics in ferromagnetic metals and their influence on ultrafast demagnetization processes. *Phys. Rev. B* **84**, 224405 (2011).

# Supplementary materials for "Demagnetization dynamics after noncollinear dual optical excitation"

Sergii Parchenko[1], Matthias Riepp[2,3], Simon Marotzke[4,5], Agne Åberg Larsson[6], Vassilios Kapaklis[6], Peter M. Oppeneer[6], Andreas Scherz[1]

1. European XFEL, Holzkoppel 4, 22869 Schenefeld, Germany
2. Université de Strasbourg, CNRS, Institut de Physique et Chimie des Matériaux de Strasbourg, UMR 7504, 67000 Strasbourg, France
3. Sorbonne Université', CNRS, Laboratoire de Chimie Physique – Matière et Rayonnement, LCPMR, 75005 Paris, France
4. Deutsches Elektronen-Synchrotron DESY, Notkestr. 85, 22607 Hamburg, Germany
5. Institut für Experimentelle und Angewandte Physik, Christian-Albrechts-Universität zu Kiel, 24098 Kiel, Germany
6. Department of Physics and Astronomy, Uppsala University, Box 516, SE-75120 Uppsala, Sweden

**Supplementary note 1. Simulation of magnetization dynamics with periodic excitation pattern**

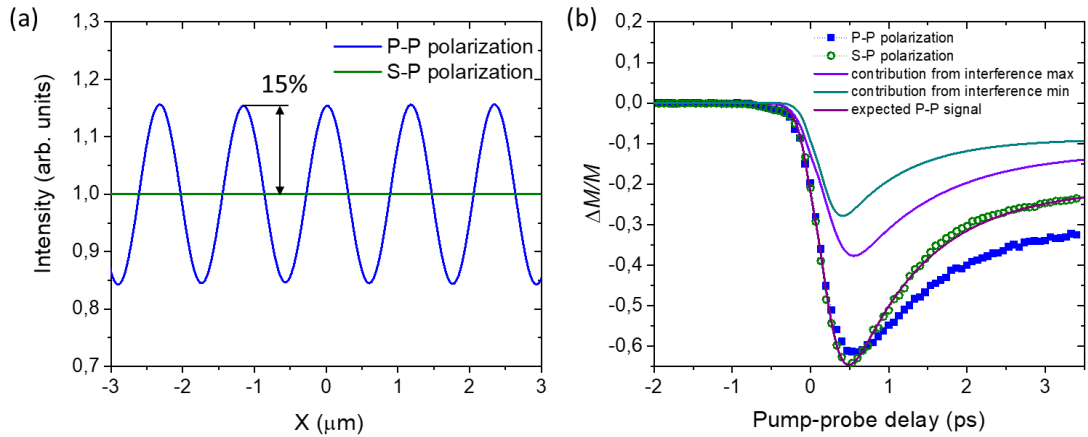

*Figure S1. (a) Normalized excitation efficiency during dual-pump experiments when the two pump pulses interfere (P-P polarization) and in the S-P configuration as a function of spatial coordinate. (b) Experimental data after dual-pump excitation and simulated tr-MORE signal after excitation in the P-P configuration and contributions from the area of the sample under interference maximal and minimal intensity.*

Because of interference between pump 1 and pump 2 in the P-P polarization configuration, we face a periodic excitation pattern. A sample is more strongly pumped when the interference is constructive and less pumped when the interference is destructive. For an angle between two pump pulses of 58 deg and a wavelength of 1030 nm, the period of the pattern is 1.13 µm, which is compatible with the wavelength of light. Thus, the intensity of light at the interference maxima or minima is only 15% larger or smaller comparable to not interferometric excitation (see Fig. S2a). Considering the linear dependence of the parameters that describe magnetization dynamics (see Fig. 1), we can simulate the magneto-optical signal considering only a periodic excitation pattern. For simplicity, we consider that half of the probed area is excited with a 15% larger fluence, and another half is excited with a 15% weaker fluence. We find that the expected signal in the P-P configuration is very close to that of the S-P configuration.

**Supplementary note 2. Single pump dynamics in TbCo**

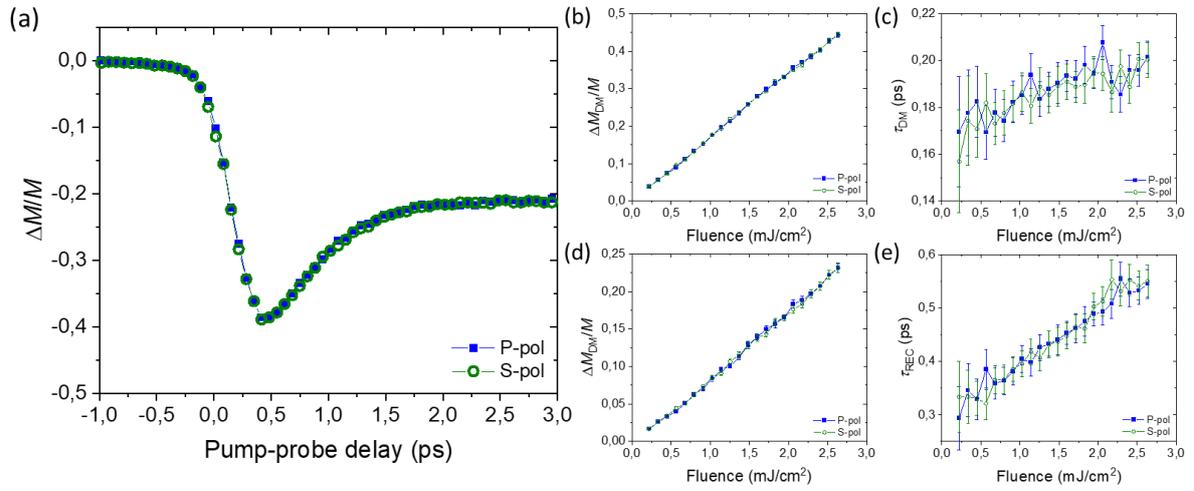

*Figure S2. (a) Time-resolved demagnetization dynamics in TbCo after excitation only with S- or P-polarized pump 1 with fluence F = 2.5 mJ/cm². The red line is fit to eq. 1. (b) Demagnetization amplitude ΔM$_{dm}$, (c) demagnetization time τ$_{dm}$ (d) recovery amplitude ΔM$_{rec}$, and (d) recovery time τ$_{rec}$ as a function of pump 1 fluence for P and S polarization.*

## Supplementary note 3. Fit results of dual pump magnetization dynamics in TbCo

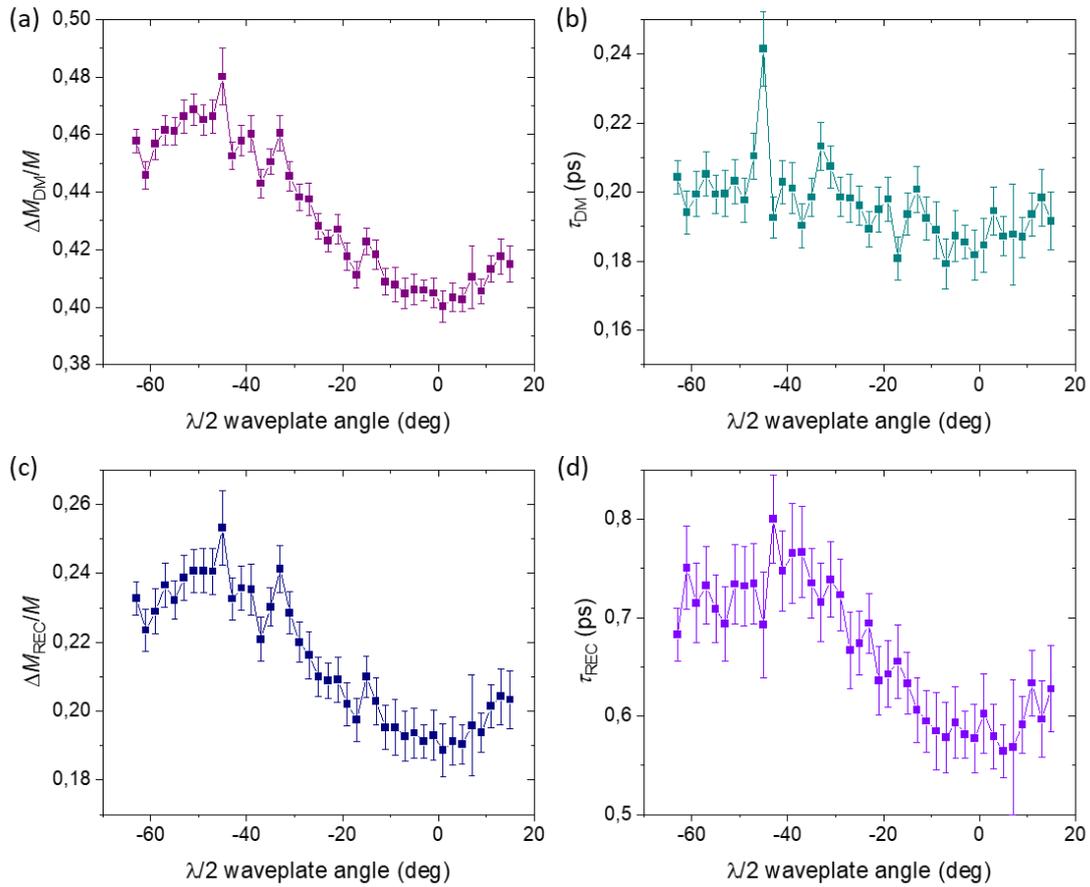

*Figure S3. (a) Demagnetization amplitude ΔM$_{dm}$, (b) demagnetization time τ$_{dm}$, (c) recovery amplitude ΔM$_{rec}$ and (d) recovery time τ$_{rec}$ in TbCo alloy dual pump excitation as a function of the pump 1 polarization direction, determined by rotation of the half-wave plate, after excitation with two pump beams. Pump 2 was always P-polarized. The fluence for the two pumps was 1.5 mJ/cm$^2$.*